\documentstyle[12pt]{article}
\newcommand{\bm}[1]{\mbox{\boldmath$#1$}}
\begin{document}
\title{A gapless charge mode induced \\  
by  the boundary states  \\
in the half-filled Hubbard open-chain}
\author{ Tetsuo Deguchi$^{1}$
, Ruihong Yue$^{2}$
  and  Koichi Kusakabe$^{\dag}$}    
\footnotetext[1]
{Institute for Theoretical Physics, 
State University of New York, 
Stony Brook, NY 11794-3840, USA, deguchi@insti.physics.sunysb.edu; 
On leave from Ochanomizu University, deguchi@phys.ocha.ac.jp . }
\footnotetext[2]
{Address after June 1998: Institute of Modern Physics, 
Northwest University,  
Xian 710069, China, yue@phy.nwu.edu.cn .}
\date{}
\maketitle
\begin{center} 
 Department of Physics, Faculty of Science \\ 
       and  Graduate  School  of Humanities and Sciences  \\  
      Ochanomizu University \\
         2-1-1 Ohtsuka, Bunkyo-Ku, Tokyo 112-8610, Japan 
\end{center} 
\begin{center} 
$^{\dag}$ Institute for Solid State Physics, \\ 
University of Tokyo \\
Roppongi, Tokyo 106-8666, Japan  
\end{center} 
\begin{abstract}
We discuss the ground state and 
some excited states of the half-filled  Hubbard model 
defined on an open chain with $L$ sites, 
where only one of the boundary sites has a 
different value of chemical potential.  
We  consider the case when the boundary site has  a negative   
 chemical potential $-p$ and the Hubbard coupling $U$ is positive.    
By an analytic method 
we show that when $p$ is larger than the transfer integral 
 some of the ground-state solutions of the Bethe ansatz equations 
become complex-valued. It follows that there 
is a ``surface phase transition'' at some critical value $p_c$; 
when $p<p_c$ all the charge excitations have the gap for the half-filled band, 
while there exists a massless charge mode when $p>p_c$.  
\end{abstract} 
%
%
%

\newpage 
The Mott-insulator transition is a fundamental phenomenon 
where the strong correlation among electrons plays an essential role. 
The existence of the insulating phase, which we call the Mott insulator, 
can not be explained within the standard framework of the  band theory.  
For the 1D Hubbard model, 
it is well known  that  under the periodic boundary condition, 
the charge gap exists only for the positive Hubbard coupling $U>0$ 
and at half-filling.  \cite{LW}
Near the transition point, however, the  system shows 
quite nontrivial many-body effects.   
\cite{UKO,KY,S-PRL,IFR}
For instance,  the effective mass diverges at half-filling 
 for the  Hubbard ring. \cite{UKO}

In order to investigate  many-body effects near the 
 transition point very precisely, let us consider 
a Hubbard chain in which only one site 
has a different chemical potential. 
With the local chemical potential we can effectively change 
the number of electrons (or holes) 
of the Mott insulator, {\it infinitesimally}.  
Let us assume  $L$ electrons in the Hubbard chain with $L$ sites.  
The system is divided into two parts; 
 a ``surface" part consisting 
of only the site with the local chemical potential,  
and a ``bulk" part of the other $L-1$ sites.
When the local potential is zero, the number of electrons in the bulk part is 
given by $L-1$; when it is very large,  no electron should occupy the 
surface site and hence all the electrons should be  in the bulk part. 
Thus, by controlling the parameter, the effective number of electrons 
in the bulk part ($L-1$ sites) can be changed continuously from $L-1$ to $L$. 
The property 
of the electrons in the bulk part 
is unique: if we consider a standard closed system, 
 the electron number will be  
 given by some integer and can not increase or decrease  
infinitesimally.

In this paper, we consider the Hubbard system defined on an open chain, 
where one of the two boundary sites is chosen as the surface site. 
We  discuss  how the half-filled ground state changes  
under the local chemical potential.      
We derive complex ground-state solutions of the Bethe ansatz equations 
by an analytic approach.  We find them 
explicitly for some finite-size systems,  
solving the Bethe ansatz equations numerically.   
Then, we calculate the energy of the ground state with the complex solutions,  
 analytically.   
It is our hope that the study of this paper might 
shed some light on some new aspects of 
the many-body effects of the  Hubbard system 
near the metal-insulator transition.

The study of this paper could be related to some real 1D systems 
such as Cu-O chain \cite{Imada}
and  quantum or atomic wires \cite{Exp,NTT}.   
The open-boundary 1D Hubbard system with the boundary 
chemical potential  could be realized in  some 1D Hubbard  system in reality, 
where the local chemical potential may play the role of a nonmagnetic 
impurity or a bias potential.

Let us introduce the 1D Hubbard Hamiltonian under the open-boundary condition,  
in which only the 1st site has the local chemical potential $-p$.  
\begin{equation}
{\cal H}
= \displaystyle - t \sum_{j=1}^{L-1}\sum_{\sigma=\uparrow,\downarrow}
      \left( c_{j\sigma}^{\dagger}c_{j+1\sigma}
       +c_{j+1\sigma}^{\dagger}c_{j\sigma}\right)
       +U \sum_{j=1}^Ln_{j\uparrow}n_{j\downarrow}
   +p \sum_{\sigma=\uparrow,\downarrow}n_{1\sigma}.
\label{Ham} 
\end{equation}  
Here  $c_{j,\sigma}$ and $n_{j,\sigma}$ stand for
the annihilation  and number operators of electron located at 
the $j$th site with spin $ \sigma$, respectively.  We recall that  
$U$ denotes the Hubbard interaction 
and $t$  the transfer integral. 
Hereafter we set $t=1$.  
The Bethe ansatz equations for 
the 1D Hubbard model have been discussed 
under some different cases of open boundary conditions. 
\cite{Schulz,AS,DY,SW}   (See also Ref. \cite{YD})
In this paper we discuss  
the open-boundary Hubbard system with $p \ge 0$.

 For $N$ electrons with $M$ down spins, 
the roots of the Bethe ansatz equations are given by 
 momenta (charge rapidities) $k_j$ 
for $j$=1 to $N$ and  rapidities (spin rapidities) $v_m$ 
for $m=1$ to $M$.   With some functions $Z_L^c(k)$ and $Z_L^s(v)$, 
the Bethe ansatz equations can be written as 
\begin{equation} 
 Z_L^c (k_j) = I_j /L \quad  {\rm for} \quad  j=1, \ldots N, \qquad  
Z_L^s(v_m) = J_m/L \quad {\rm for} \quad v=1, \ldots, M.  
\label{BAE} 
\end{equation}
Here the quantum numbers $I_j$ and $J_m$ are given by some integers.

Let us consider the half-filled band 
under the boundary chemical potential,   
where $N=L$ and $M=L/2$.    
Hereafter we  assume that $L$ is even. 
We consider  analytic continuations of 
the functions $Z_L^c(k)$ and $Z_L^s(v)$ with respect to the parameter $p$. 
Let us introduce  an adiabatic hypothesis that 
the quantum numbers $I_j$ and $J_m$ should be constant when we continuously 
change the parameter $p$. 
Under the hypothesis, all the solutions of the Bethe ansatz equations 
can be labelled by their quantum numbers. 
When $p=0$, we can order the ground-state roots $k_j$'s and $v_m$'s  
such that $I_j=j$ for $1 \le j \le L$ and $J_m=m$ for $1 \le m \le L/2$.  
The hypothesis is consistent with our analytic arguments 
and numerical results. 
Thus, for any value of $p$, 
the quantum numbers of momentum $k_j$ and rapidity $v_m$ are given by 
$j$ and $m$, respectively.

We now consider the Bethe ansatz equations more explicitly.  
Let us denote by $I_{max}$ ($I_{min}$) the largest (smallest) 
integer of  the quantum numbers of real momenta over all possible excitations  
and by $J_{max}$ ($J_{min}$)  that of 
real rapidities. Then,  the set $\Delta_{re}^c$  ($\Delta_{re}^s$) 
of all the possible quantum numbers $I_j$'s ($J_m$'s)
 for real momenta (rapidities) are given by 
\begin{equation}
\Delta_{re}^c = \{I_{min}, I_{min}+1, \ldots, I_{max} \}  \quad 
\left(\Delta_{re}^s = \{J_{min}, J_{min}+1, \ldots, J_{max} \} \right).  
\end{equation}
Let us write by $\Delta_{hole}^c$ ($\Delta_{hole}^s$) the set of 
the quantum numbers  of holes 
of real momenta (rapidities) in the ground state. 
Then, the set of the quantum numbers of the real momenta (rapidities)  
for the  ground state is given by 
$\Delta^c_g=\Delta_{re}^c - \Delta_{hole}^c$ ($\Delta^s_g=\Delta_{re}^s-
\Delta_{hole}^s$). 
Let us denote by $I_{max}^g$ ($J_{max}^g$) the largest integer of the set 
$\Delta_g^c$ ($\Delta_g^s$). Then, $I_{max}^g \le I_{max}$, in general.   
We introduce the symbol 
$\Delta_{im}^c$ ($\Delta_{im}^s$) for the set of the quantum 
numbers for complex-valued momenta (rapidities) in the ground state.    
In terms of the symbols, 
the functions $Z_L^c(k)$ and $Z_L^s(v)$ for the ground state 
are written as follows
\begin{equation}
\begin{array}{rcl}
\displaystyle Z_{L}^c(k)&=&\displaystyle \frac{2k}{2\pi}+ 
  \frac1{ L}\sum_{n \in \Delta^s_g} \sum_{r=\pm1} 
\theta_1(\sin k-r v_n)
+{\frac 1  L} z_B^c(k) , 
\\[3mm]
\displaystyle Z_{L}^s(v)&=&
\displaystyle 
 \frac1{ L}\sum_{j \in \Delta^c_g } \sum_{r=\pm1} \theta_1(v- r \sin k_j)
  -\frac1{ L}\sum_{n \in \Delta^s_{g}} \sum_{r=\pm1} \theta_2(v-r v_n)
+ {\frac 1 L} z_B^s(v) , 
\end{array}
\label{ZLcs}
\end{equation}
where the functions $z_B^c(k)$ and  $z_B^s(v)$ are given by  
\begin{equation}
\begin{array}{rcl} 
\displaystyle  z_B^c(k)& =& 
 \displaystyle {\frac {2k}{2 \pi}} - {\frac 1{ 2\pi i}}
  \log\left(\frac{1+ p e^{ik}}{1+ pe^{-ik} }\right)
  + \sum_{m \in \Delta_{im}^s} \sum_{r=\pm1} \theta_1(\sin k-r v_m),  
  \\[3mm]
\displaystyle  z_B^s(v) &= &  \displaystyle\theta_1(v)
  +\sum_{j \in \Delta_{im}^c} \sum_{r=\pm1} \theta_1(v - r \sin k_j)
  -\sum_{m \in \Delta_{im}^s } \sum_{r=\pm1} \theta_2(v-r v_m) . 
\\[3mm]
\end{array} . 
\label{zBcs}
\end{equation}
Here, the functions $\theta_n(x)$ have been defined by 
$\theta_n(x)=2\tan^{-1}\left( x/(nu)\right)/(2\pi) $,
where $u$ is given  by $u=U/4$.   
An outline of the derivation of the Bethe ansatz equations 
is given in Appendix A. 
%
%

When $p$ is larger than some critical values of $p$, 
some of the  ground-state solutions become complex-valued.   
The number of complex roots is different 
for  four regions of $p$, which are divided by the 
 critical values $p_{cj}$'s.  They are 
given by 
$p_{c1} = 1$, 
 $p_{c2} = u + \sqrt{ 1+ u^2}$, 
$p_{c3} = 2u + \sqrt{ 1+ 4 u^2} $. 
Let us introduce some notation. 
We define symbol $\kappa$ 
by  $\kappa=\log |p|$ for $p>0$ and $p<0$.  
We also define $\alpha$ by  
$\alpha = \sinh \kappa /u$ for $p>0$ and $p<0$.  
The notation of the 
critical points is summarized as 
$p_{cj}=(j-1)u+ \sqrt{1+ (j-1)^2u^2}$ for $j=1,2,3$. 
If a set $\Delta$ is empty, we denote it by 
$\Delta = \phi$. 
Then, the sets of quantum 
numbers are given by the following. 
\begin{enumerate}
\item For $0 < p < p_{c1} $,
we have no boundary solutions. 
The sets of quantum numbers are given by 
$$
\Delta_g^c = \{ 1,2,\ldots, L\} , \qquad \Delta_g^s= \{1,2,\ldots, L/2\}, 
$$
$$
\Delta_{im}^c  = 
\Delta_{im}^s=  
\Delta_{hole}^c= 
\Delta_{hole}^s= \phi . 
$$
The $I_{min}$'s are given by the following 
$$
I_{min} = 1, \quad 
I_{max}=L, \quad 
J_{min}=1, \quad 
J_{max}=L/2.  
$$
\item  For $p_{c1}< p< p_{c2}$ ($0 < \alpha < 1$), 
we have a complex-valued momentum $k_L$ given by   
\begin{equation}
k_L= \pi + i \kappa - i \delta_L . 
\end{equation}
The sets of quantum numbers are given by 
$$
\Delta_g^c = \{2,\ldots, L\} , \qquad \Delta_g^s= \{1,2,\ldots, L/2\}, 
$$
$$
\Delta_{im}^c=\{ 1 \}, \quad 
\Delta_{im}^s=
\Delta_{hole}^c=
\Delta_{hole}^s= \phi . 
$$
The $I_{min}$'s are given by the following  
$$
I_{min}=1, \quad 
I_{max}=L-1, \quad 
J_{min}=1, \quad 
J_{max}=L/2 . 
$$
\item For $p_{c2} < p< p_{c3}$ ($1 < \alpha < 2$), 
we have the complex momentum $k_L$ and the complex rapidity $v_1$ given by 
\begin{eqnarray}
k_L & =  & \pi + i \kappa -i \delta_L, \nonumber \\ 
v_1 & = & i (\alpha -1) u + i \eta_1 . 
\end{eqnarray} 
The sets of quantum numbers are given by 
$$
\Delta_g^c = \{2,\ldots, L\} , 
\qquad \Delta_g^s= \{2,\ldots, L/2\}, 
$$
$$
 \Delta_{im}^c=\{ L \}, \quad 
 \Delta_{im}^s=\{1 \}, \quad 
 \Delta_{hole}^c= \Delta_{hole}^s=\phi . 
$$ 
The $I_{min}$'s are given by the following. 
$$
I_{min}=1, \quad 
I_{max}=L-1, \quad 
J_{min}=2, \quad 
J_{max}=L/2 .
$$
\item For $p_{c3} < p $ ($2 < \alpha $), 
we have the following three complex roots $k_1$, $k_L$ and $v_1$ 
\begin{equation}
\begin{array}{rcl}
k_{1} & = & i \log \left( (\alpha -2)u + 
\sqrt{(\alpha - 2)^2 u^2 +1} \right) + i \delta_1,  \\
k_L & =& \pi + i \kappa -i \delta_L, \\ 
v_1 & = & i (\alpha -1) u + i \eta_1 .
\end{array}
\label{bkL}
\end{equation}     
We call  them a {\it boundary $k-\Lambda$ string}. 
The sets of quantum numbers are given by 
$$
\Delta_g^c = \{2,3, \dots, L-1\}, 
\qquad \Delta_g^s = \{ 2,3, \ldots, L/2 \}, 
$$
$$
\Delta_{im}^c=\{1, L \}, \quad 
 \Delta_{im}^s=\{1\}, \quad 
 \Delta_{hole}^c=\{L\}, \quad 
\Delta_{hole}^s=\phi . 
$$ 
The $I_{min}$'s are given by the following  
$$
I_{min}=2, \quad 
I_{max}=L, \quad 
J_{min}=2, \quad J_{max}=L/2 . 
$$  
\end{enumerate}
We note that when  $p > p_{c3}$,  a hole of real momenta appears in 
the half-filled ground state at $I=L$;   
$I_{max}^g =L-1$  and  $I_{max}=L$ when $p> p_{c3}$.  
We also note that $\delta_1, \delta_L$  and $\eta_1$ are exponentially 
small except for some neighborhoods of the critical points.     
 For instance, we can show   
$\delta_1= O(p^{-2L})$ for $p_{c1}<p<p_{c2}$.  
The quantities $\delta_L$, $\delta_1$ and $\eta_1$
are explicitly evaluated in Appendix B. 
%
%

For the case when $p<0$, some complex boundary solutions  
 have been  discussed for the 1D Hubbard model under 
the open-boundary conditions \cite{TY,BF,EF},  where the quantum numbers 
of the complex rapidities $k_1^{'}$, $k_2^{'}$ and $v_1^{'}$ 
correspond to  $I_1=1, I_2=2$ and $J_1 = 1$, respectively.
%
Furthermore, when the band-width $t$ is very large and 
the electron density $N/L$ is very small, 
the boundary solutions $k_1^{'}$, $k_2^{'}$ and $v_1^{'}$ 
for the case of $p<0$ can 
correspond to the boundary solutions of the 
1D interacting spin-1/2 Fermi system, which had been 
discussed in Ref. \cite{spin-1/2}. 
(See also  Appendix B.)

For the half-filling case, 
the ground-state energy for $p>0$ is related to 
that of $p<0$ through the particle-hole transformation, which 
will be discussed in Appendix C. For instance, 
the energy of the ground state for $p>p_{c3}$ with 
the boundary solutions $k_L$, $k_1$ and $v_1$, 
 is transformed 
into that of $p<-p_{c3}$ with 
$k_1^{'}$, $k_2^{'}$ and $v_1^{'}$. However, 
it seems quite  non-trivial how the two sets of 
the charge  rapidities 
for the two cases of $p>0$ and $p<0$ 
 could be  related to each other.    
(See also Appendix C.) 
%
%

%
%

Let us show that  momentum $k_L$ which is real-valued when $p<p_{c1}$  
becomes  complex-valued when $p > p_{c1}$. 
 First, we note that 
when $k$ is real and  $|\pi-k| \ll 1$, we have 
\begin{equation}
{\frac 1{2 \pi  i}}
  \log\left(\frac{1+ p e^{ik}}{1+ pe^{-ik} }\right)
=  {\rm H}(p-p_{c1}) + {\frac 2 {2 \pi }}
\tan^{-1}({\frac {p \sin(\pi -k)} {1 - p \cos(\pi - k)}}) . 
\label{pc1}
\end{equation}
Here ${\rm H}(x)$ denotes the Heaviside step-function: 
${\rm H}(x)=0$ for $x<0$ and 
${\rm H}(x)=1$ for $x>0$. 
Suppose that momentum $k_L$  be real 
even when $p> 1$.  Since $k_L$ is  close to $\pi$, 
we have $Z_L^c(k_L)=k_L/\pi + z_B^c(k_L)/L$. 
It follows from  (\ref{pc1}) 
 that the value of $z_B^c(k_L)$ for $p>1$ is  by  $1$ smaller than   
that of the case when $p<1$:  $z_B^c(k_L) = k_L/\pi - 1 + O(1/L)$ 
for $p>1$. Thus, we have $I_L/L =  k_L/\pi + 
(k_L/\pi-1)/L + O(1/L^2)$, which leads to $k_L= \pi + O(1/L^2)$ for 
$I_L =L$. However, when $k=\pi$ 
the wave function should vanish under the open-boundary 
condition. Thus, we arrive at an inconsistency. Therefore, 
the momentum $k_L$ should be complex-valued when $p>1$.

We can show  that  $v_1$ becomes imaginary when $p>p_{c2}$. 
Let us take  the following branch of the logarithmic function: 
 $-i \log\left(e(x) \right) =  \pi - 2 \tan^{-1}(x)$, 
where  $e(x)$ denotes $e(x)=(x+i)/(x-i)$. 
 Then, we can show   
\begin{equation}
\sum_{r=\pm 1} \theta_n(v + r i \gamma u ) = 
 \displaystyle 
\left\{ 
\begin{array}{cc} 
\theta_{\gamma+n}(v) + \theta_{n-\gamma}(v), 
& {\rm for} \quad \gamma < n ,\\ 
\theta_{\gamma+n}(v) + 1 - \theta_{\gamma-n}(v),  
& {\rm for} \quad \gamma > n .
\end{array} 
\right. \label{shift}
\end{equation}
Applying the formula (\ref{shift}) with $\gamma =\alpha$ 
to the function $z_B^s(v)$, we can show that if we assume  
the smallest rapidity $v_1$ to be real, then  it would be 
$O(1/L^2)$ for $p>p_{c2}$, and also that therefore  
it should be imaginary when $p>p_{c2}$.  
In the same way with the rapidity $v_1$, using the formula (\ref{shift})  
we can also show that momentum $k_1$  becomes imaginary  when  $p>p_{c3}$.

We can evaluate 
the largest and smallest integers of all the  possible quantum numbers for 
real momenta  in the following way. 
The function $Z_L^c(k)$ is monotonically  increasing with respect to $k$, 
since the density of real momenta should be non-negative. 
 We note that 
under the open boundary condition, the Bethe-ansatz wavefunction  
should vanish if there exists a momentum of $k=0$ or $\pi$.  
Thus, the equations for  $I_{min}$ and $I_{max}$ are given by  
\begin{equation}
Z_L^c(0)= (I_{min}-1)/L, \quad Z_L^c(\pi)= (I_{max}+1)/L .  
\label{Iminmax}
\end{equation}
We  determine $I_{min}$ and $I_{max}$ by solving eqs. (\ref{Iminmax}).  
 For instance, for the case when $0 \le p < 1$, 
it is easy to see  $Z_L^c(0)=0$ and $Z_L^c(\pi)=(L+1)/L$, so that 
we obtain  $I_{min}=1$ and $I_{max}=L$.

 For  real rapidities, we can obtain  $J_{min}$ 
and $J_{max}$ by applying the argument in Ref \cite{Takhtajan}. 
It is easy to show that they satisfy the following equations. 
\begin{equation}
Z_L^s(0)=(J_{min}-1)/L, \quad Z_L^s(\infty)=(J_{max}+1)/L.
\label{Jminmax}
\end{equation}
Solving eqs.(\ref{Jminmax}) we determine $J_{min}$ and $J_{max}$. 
 For example, let us consider the case 
$p_{c2} <  p < p_{c3}$. From eqs. (\ref{ZLcs}) and (\ref{zBcs}) 
we can show  $Z_L^s(\infty)=1+(1-J_{max})/L$. 
Thus, we obtain $J_{max}=L/2$. 
We can discuss the maximal and minimal 
quantum numbers also for some excited states with 
boundary solutions, similarly. 
Some details are given in Appendix D. 
%
%

The 
 new hole appears in the half-filled band, when $p>p_{c3}$. 
Therefore, there is 
a gapless mode of 
particle-hole excitations for 
the half-filled ground state 
under the open boundary condition.  Let us give some explanation 
in three paragraphs in the following

First, we consider  the appearance of the new hole. 
This is a consequence of the formation of the boundary $k-\Lambda$ string. 
In fact, the number of possible real momenta in the band 
is given by $L-1$, since $I_{max}=L$ and $I_{min}=2$ when $p> p_{c3}$.
On the other hand, there are only $L-2$ real momenta in the wavefunction 
since we have two complex momenta $k_1$ and $k_L$.  
Thus, there should be one hole in the band. 
By shifting the quantum number of the hole, we can make 
a series of charge excitations with the hole;  for the ground state 
the quantum number of the hole is given by $L$, 
while for the charge excited state it is given by an integer less than $L$.

Second, we consider 
the gap energy for the 
charge  excited state, 
where the quantum number of the hole is close to $L$ (for example, $L-1$). 
Then, we see that it becomes  infinitesimally small 
when we take  the thermodynamic limit $L \rightarrow \infty$.  
In this sense, we may call the mode  gapless. 
Here we note that the excitation energy should be continuous 
with respect to the charge rapidity $k_h$ of the new hole.

Furthermore, we can explicitly calculate   
the charge excitation energy,   
applying the 
method \cite{W2} of the finite-size correction. 
We recall that $k_h$ denotes the charge rapidity of the new hole.   
We denote by $E_L^{ex}(k_h)$ the energy of  
the charge excited state with the new hole.  
Then it is given by 
\begin{equation}
E_L^{ex}(k_h) = E_L^g - 2e^c(k_h) + 2 e^c(\pi),  \label{excited}
\end{equation}
where $E_L^g$ denotes  the ground-state energy 
for $p>p_{c3}$ and $e^c(k)$ is the dressed energy 
\cite{W2}   
for the half-filled band given by 
\begin{equation}
e^c(k)= - {\frac A 2} -\cos k - 
\int_{-\infty}^{\infty}  
\frac {e^{-u \omega} J_1(\omega) \cos(\omega \sin k)}
{\omega \cosh u \omega } d \omega . 
\end{equation}
The expression of the chemical potential $A$ 
at the half-filling will be given later in (\ref{A}). 
\setcounter{footnote}{1}
\footnote{
The expression (\ref{excited}) can be derived from the formula 
(\ref{formula}) by replacing  the hole momentum $k_h^g$ 
of the ground state 
by that of the  excited state.} 
 From the expression of the excited energy (\ref{excited})
we see that the gap energy of the mode 
is of the order of $1/L^2$. Thus, we see that 
the gap energy vanishes 
under the thermodynamic limit:  
$L \rightarrow \infty$. 


The three boundary complex solutions for $p>p_{c3}$ can be considered as 
a variant of $k-\Lambda$ string that was originally defined for the periodic 
Hubbard model.  
In fact,  we can derive the expressions (\ref{bkL}) of 
the boundary $k-\Lambda$ string from the viewpoint of 
classification of $k-\Lambda$ strings of length $n=1$.  
Details will be discussed in later papers.

Let us explicitly study for a finite-size system 
the behaviors of momenta and rapidities 
with respect to the boundary chemical potential. 
In Fig. 1, the flows of momenta and rapidities are plotted 
versus the parameter $p$ 
for the 8-sited Hubbard Hamiltonian under the open boundary 
condition, where the roots are 
obtained numerically 
by solving the Bethe ansatz equations with $L=N=8$ and $M=4$. 
As far as the finite-size systems we have investigated are concerned, 
the numerical solutions are consistent with 
the following consequences of the analytic approach: 
the complex solutions are formed one-by-one at the critical points 
of the parameter $p$; there is a charge hole when $p>p_{c3}$. 
This is nontrivial. The analytic method should be valid only 
when the system size is very large.    However, 
these important properties are already observed   
in such a small system as the case of $L=8$. 
$$
Fig. 1 
$$

Let us explain how we apply to our system 
 the method \cite{W2}  of the finite-size correction. 
We consider the Hamiltonian ${{\cal H}^{'}} = {\cal H} - A N - h (N-2M)/2$, 
where $A$ and $h$ are the chemical potential 
and the uniform magnetic field, respectively. 
In order to define densities of the roots of the Bethe ansatz equations,  
 we extend  $Z_L^c(k)$ and $Z_L^s(v)$ 
into odd functions defined both on positive and negative values 
of their variables. 
 For an illustration, we consider the density  of real-valued rapidities. 
When $0 \le p < p_{c2}$, we have  
$Z_L^s(0)=0$ and the function $Z_L^s(v)$ itself can be simply  
extended into an odd function of $v$ by  $Z_L^s(-v) = -Z_L^s(v)$ for $v > 0$.  
We define  rapidity with negative suffix by  $v_{-m}=-v_m$ for 
$m=1, \ldots , L/2$.  Then, the density of the real rapidities is given by 
the derivative $\rho^s_L(v)=dZ_L^s(v)/dv$ for $-\infty < v < \infty$. 
When $p_{c2} < p$, however, the function does not vanish 
at the origin: $Z_L^s(0)=1/L$. In this case, we  introduce 
some shifts of the function and the variable  
${\tilde Z}_L^s(v)=Z_L^s(v)-1/L$   and ${\tilde v}_m = v_{m+1}$, 
respectively. We also introduce rapidity of negative suffix by  
${\tilde v}_{-m}= {\tilde v}_m$, for $m>0$. 
Then, the Bethe ansatz equations are given by 
\begin{equation}
{\tilde Z}_L^s({\tilde v}_m)={\tilde J}_m/L, \qquad  {\rm for } 
\quad m=-{\tilde J}_{max}^g, \ldots, {\tilde J}_{max}^g,
\end{equation}
where ${\tilde J_m}=m$ and ${\tilde J}_{max}^g=J_{max}^g-1$.  
Then, we can safely define   the density of rapidities by the derivative 
$\rho_L^s({\tilde v})=d{\tilde Z}_L^s({\tilde v})/d{\tilde v}$.

Taking the derivatives of the Bethe ansatz 
equations together with some  
 continuous limits, 
we can systematically derive a set of equations 
for the densities of the system with $L$ sites. 
 For the half-filled band under zero magnetic field, 
the set of equations for  the densities up to $O(1/L)$ 
is given  in the following 
\begin{equation}
\label{BAE2}
\begin{array}{rcl}
\rho_L^c(k)&=&\displaystyle {\frac1{\pi}}+{\frac{1}{L}}\tau^{c (0)}(k)
  + \cos k \int_{-\infty}^{\infty}a_1(\sin k- v) 
\rho^s_L(v) dv + O(1/L^2), \\
\rho^s_L(v)&=&\displaystyle {\frac{1}{L}}\tau^{s(0)}(v)
            +\int_{-\pi}^{\pi}a_1(v-\sin k ) \rho^c_L(k)dk \\ 
     &  & \displaystyle
  \qquad  - \int_{-\infty}^{\infty}a_2(v- v')\rho^s_L(v') dv' 
+ O(1/L^2).  
\end{array} 
\label{BAEdensity}
\end{equation}
Here $a_n(x)$ is defined by  $2\pi a_n(x)=(2nu)/(x^2+(nu)^2)$.   
The boundary terms $\tau^{c(0)}(k)$ and $\tau^{s(0)}(v)$ in eqs. 
(\ref{BAEdensity}) 
are given by the derivatives 
of $P_0(k)/(2\pi)$ and $Q_0(v)/(2\pi)$, respectively,  
where they are related to $z_B^c(k)$ and $z_B^s(v)$ by  
\begin{equation}
P_0(k)/2\pi= z_B^c(k) - \theta_1(\sin k), \quad  
Q_0(v)/2\pi= z_B^s(v) - \theta_1(v) + \theta_2(v)  . 
\end{equation}

We now  evaluate the ground-state energy $E_L^g$ of 
the Hamiltonian ${\cal H}^{'}$ 
at half-filling under zero magnetic field. From eqs. 
(\ref{BAEdensity})  we have 
the following 
\begin{eqnarray}
E_L^g & = & - \sum_{j \in \Delta_g^c}  
2\cos k_j - \sum_{ j \in \Delta_{im}^c} 2 \cos k_j -A N  
\nonumber \\
& = & L e_{\infty} + 1+ A/2 + ({\bm e},{\bm \tau}^{(0)}) 
- \sum_{h \in \Delta^c_{hole}} 2e^c(k_h^g) \nonumber  \\
& & - \sum_{j \in \Delta_{im}^c}(A + 2 \cos k_j) +  O(1/L) ,   
\label{formula} 
\end{eqnarray}
where ${\bm \tau}^{(0)}(k,v) = (\tau^{c(0)}(k), \tau^{s(0)}(v))$ denotes the 
surface density,  the symbol
${\bm e}=(e^c(k),e^s(v))$ denotes the dressed energy. \cite{W2,DY}   
We recall that $k_h^g$'s denote the momenta of  
possible holes at the ground state. 
(For the ground state of $p>p_{c3}$, we have only one hole. )
The inner product $({\bm e}, {\bm \tau}^{(0)})$ is defined 
by the following \cite{W2,DY}
\begin{equation}
({\bm e}, {\bm \tau}^{(0)}) = 
\int_{-\pi}^{\pi} e^c(k) \tau^{c(0)}(k) dk +
 \int_{-\infty}^{\infty} e^s(v) \tau^{s(0)}(v) dv  . 
\end{equation}
Let us define the surface energy $e_{sur}$ of the system by 
the $O(1)$ part of the ground-state energy. Then it is given 
in the following  
\begin{enumerate} 
\item 
 For $0< p< 1$, 
\begin{equation}
e_{con}+ p- \sum_{n=0}^{\infty} p^{2n} \int_{0}^{\infty} 
{\frac  {2 e^{-u \omega} J_1(\omega) 
J_{2n}(\omega)}{\omega \cosh u \omega}} d\omega .  
\label{energy1}
\end{equation}
\item 
 For $1< p< p_{c3}$, 
\begin{equation}
e_{con} + p- \int_0^{\infty} 
{\frac {2e^{-u\omega} \cosh(\omega \sinh \kappa ) J_1(\omega)} 
{\omega \cosh u \omega}}d \omega
+ \sum_{n=1}^{\infty}{\frac 1  {p^{2n}}}  \int_{0}^{\infty} 
{\frac  {2 e^{-u \omega} J_1(\omega) 
J_{2n}(\omega)}{\omega \cosh u \omega}} d\omega  . 
\label{energy2}
\end{equation}
\item 
 For $p_{c3} < p$, 
\begin{equation}
e_{con}+ 4u -A - {\frac 1 p} 
+ \sum_{n=1}^{\infty} {\frac 1 {p^{2n}}} \int_{0}^{\infty} 
{\frac  {2 e^{-u \omega} J_1(\omega) 
J_{2n}(\omega)}{\omega \cosh u \omega}} d\omega .    
\label{energy3}
\end{equation}
\end{enumerate} 
Here the symbol $e_{con}$ denotes the surface energy for $p=0$, which is 
explicitly given by 
\begin{equation}
e_{con}=
(1-A/2)  + 2\sqrt{1 +u^2} -2u - \int_0^{\infty} 
{\frac {e^{-2u\omega}J_1(\omega)} {\omega \cosh u \omega}}d \omega .   
\end{equation} 
The chemical potential $A$ at half-filling is given by 
\begin{equation}
A= 2 - 2 \int_0^{\infty} 
{\frac {e^{-u\omega}J_1(\omega)} 
{\omega \cosh u \omega}}d \omega . \label{A}  
\end{equation}

Let us discuss the ground-state energy for the strong-coupling case. 
When $p>p_{c3}$, it becomes close 
to the energy of the first charge-excited state for $p=0$. 
We compare the surface energy 
 for $p=0$ given by eq. (\ref{energy1}) with that of  $p>p_{c3}$ 
given by eq. (\ref{energy3}).  
Then, the main part of the difference between them 
is  given by $4u=U$, which is almost equivalent 
to the charge-gap energy $4u-2A$ at $p=0$. 
We note that when $u \gg 1$, we have  $p_{c3} \gg 1$ and $u \gg A$.

Under the strong coupling condition, the main part of the surface energy 
is given by the following; $p+e_{con}$ when $1 \ll p < p_{c3}$ and 
$4u-2A + e_{con}$ when $p > p_{c3}$. 
>From the calculation of the ground-state energy,  we can evaluate  
the average number $n_1$ of electrons on 
the 1st site, since it is defined  by $n_1={\partial E_L}/{\partial p}$.  
We find that  $ {\partial E_L}/{\partial p} \approx 1$ for $1 \ll p < p_{c3}$ and  
$ {\partial E_L}/{\partial p} \approx 0$ for $p_{c3} < p $.  This suggests 
that one hole should be localized at the surface site 
when $p>p_{c3}$.  
The result is consistent with the  discussion over the complex boundary solutions   
that the half-filled ground state has  gapless charge-excitations  
when $p>p_{c3}$ since one hole appears in the band.

Let us discuss  the spectrum of a finite-size system numerically.     
The low-excited spectrum of the 6-sited open Hubbard Hamiltonian 
with $U=20t$ is obtained by    
the exact numerical diagonalization of the Householder method.    
The spectral flows with respect to the parameter $p$ 
are depicted in  Fig. 2. 
$$
Fig. 2
$$
 From Fig. 2 we see  that the energy levels of 
charge excitations become  close to   
the ground-state energy at $p=p_{c3}$.  

 From  Figs. 1 and 2, we have  the 
following observations, respectively.  (i) When $p> p_{c3}$ 
the first charge-excited state can be obtained by 
shifting the position of the hole in the band of real momenta $k_j$'s; 
such shifting is equivalent to taking a different quantum number for the hole.  
(ii) The energy level of the first charge-excited state for $p> p_{c3}$ 
is identified with that of the lowest state above the charge gap for $p=0$;  
 we can trace the spectral flow of the excited state 
from $p=p_{c3}$ down to $p=0$ in Fig. 2. 
>From the analytic approach, the observation (ii) should hold  due to   
 the adiabatic hypothesis on the quantum numbers.  
 From  (i) and (ii), we can say that 
the characteristic properties of the energy spectrum discussed   
by the analytic method  
are also in common with that of the finite-size system.    
Thus, the spectrum of the 6-sited system may illustrate that of  
thermodynamically large systems.

In this paper we have discussed the boundary solutions 
for the half-filled band when $p>0$.  
We have  shown that when $p>p_{c3}$ 
one mode of charge excitations has the gap energy 
of the order of $1/L^2$; we call it massless since the gap vanishes 
in the thermodynamic limit. 
 By the method of  the finite-size correction, 
we have   calculated  the ground-state energy and 
the excited energy of the massless mode.  
We note that  it is not difficult to derive  
explicit formulas  for the energies of  other excitations.  
In fact, all the spectrum shown in Fig. 2 can be explained analytically. 
Details will be given in the next paper. \cite{YDK}

\vskip 0.6cm 
\noindent 
{\bf Acknowledgements}  

Two of the authors (T.D. and K.K)  would like to thank N. Kawakami and 
H. Tsunetsugu for useful comments. 
R.Y.  was granted by the JSPS foundation 
and the Monbusho Grant-in-Aid of Japanese Government. 
One of the author (T.D.) would like to thank 
V. Korepin for useful comments.

\newpage
{\bf Figure Captions}
\vskip 1.2cm 
\begin{flushleft}
fig.1 
\end{flushleft}

\par \noindent 
(a) Flow of momenta (or charge rapidities) as a function of boundary potential ($p$)
      for the one-dimensional open Hubbard model with 8 sites
      at the half filling ($N=L$) and $U=20t$.
      Solid lines represent real momenta.
      At $p=p_{c1}\sim t$, the largest charge rapidity approaches $\pi$
      and becomes complex, $\pi + {\rm i} \kappa$, whose complex part 
      is given by the dashed line, for $p>p_{c1}$.
      Beyond $p=p_{c3}\sim U$, the smallest momentum becomes complex, 
      $-{\rm i} \kappa'$, where $\kappa'$ is represented by 
      the dot-dashed line.
      (b) Flow of rapidities (or spin rapidities) for the same system. 
      Solid lines represent real rapidities.
      At $p=p_{c2}\sim U/2$, the smallest rapidity becomes complex, 
      ${\rm i} \chi$, where $\chi$ is given by the dashed line.
\begin{flushleft}
Fig.2 
\end{flushleft}
\par \noindent 
Spectral flow for the 6-site open Hubbard chain at the half-filling 
      with $U=20t$ as a function of  boundary potential $p$.
      Dots denote all of the eigenvalues for this system obtained by 
      the direct diagonalization. 
      The lower solid line represents the ground state energy given by 
      the Bethe ansatz. The upper solid line corresponds to the first charge-excited 
      state, which can be traced back from $p=p_{c3}$ to $p=0$;   
      at $p=0$, it is the lowest level beyond the charge gap. 
      Enlarged flow around the gap-closing transition point is
      depicted in the inset. 
      Below a critical point ($p_{c3}$), a charge gap exists 
      above the continuum of low-energy spin excitations.

\newpage

\setcounter{equation}{0} 
\renewcommand{\theequation}{A.\arabic{equation}}
\section{Appendix A:}

Let us briefly outline  the  derivation of 
the Bethe-ansatz equations through the  algebraic Bethe-ansatz 
method for the open-boundary XXZ model given by E.K. Sklyanin.  
Some details can be found in Ref. \cite{DY}. (See also \cite{YD}.) 
We write the eigenstates   
for $N$ electrons with $M$ down-spins as 
\begin{equation}
\Psi_{NM}=\sum f_{\sigma_1, \ldots ,\sigma_N} 
(x_1,\cdots,x_N)c^{\dagger}_{x_1\sigma_1}\cdots
c^{\dagger}_{x_N\sigma_N}|vac \rangle . 
\end{equation}
Here,  $x_j$ and  $\sigma_j$ are 
the position and spin variables of the electrons, respectively. 
In the region $x_{Q1}\leq\cdots\leq x_{QN}$, 
we assume that the Bethe-ansatz wavefunction $f$
takes the form 
\begin{equation}
f_{\sigma_1, \ldots, \sigma_N} 
(x_1,\cdots,x_N)=\sum_{P}\epsilon_P
   A_{\sigma_{Q1},\cdots,\sigma_{QN}}
(k_{P1},\cdots,k_{PN})
\exp\{i\sum_{j=1}^N k_{Pj}x_{Qj}\}. 
\end{equation}
Here the $Q$ is an element of $S_N$, the permutation group of $N$
particles, and $P$ runs over all the permutations 
and the ways of negations of $k's$; 
there are $N ! \times 2^N$ possibilities for $P$, while $N!$ for $Q$.  
We employ the notation: $k_{-j}=-k_j$. 
The symbol  $\epsilon_P$ denotes the sign of $P$; 
if the permutation is even, $P$ makes $\epsilon_P=-1$ when odd number of 
$k$'s are negative and $\epsilon_P=1$ when even number of $k$'s are negative.
Let us introduce the vector 
${\vec A}(k_{j_1}, \ldots, k_{j_N})$ such that  
its element for  entry $(Q1,\ldots, QN)$ 
is given by 
$A_{\sigma_{Q1},\cdots,\sigma_{QN}}
(k_{j_1},\cdots,k_{j_N})$. 
Here we note that the suffix 
$j_1, \ldots, j_N$ can be written as  
 $P1, \ldots, PN$, respectively,  by some $P$. 
Then, we can show that the consistency condition 
for the amplitudes  ${\vec A}(k_{j_1},\ldots ,k_{j_N})$   
is given by the following   
\begin{equation}
T(\sin k_{P1}) {\vec A}(k_{P1},\dots ,k_{PN})
={\vec A}(k_{P1},\dots ,k_{PN}) . 
\end{equation}
Here $T(u)$ is the inhomogeneous transfer matrix of 
the open-boundary XXX model with $N$ inhomogeneous parameters 
$\sin k_{P1}$, \ldots, $\sin k_{PN}$. \cite{DY}
Let us denote the eigenvalue of the transfer matrix $T(u)$ by 
$\Lambda(u)$. Then, from the condition 
$\Lambda(\sin k_{P1})=1$ and the 
 Bethe ansatz equations for the XXX model, 
 the Bethe ansatz equations of the 1D open-boundary 
 Hubbard model for the charge and spin parts are 
 obtained, respectively.

 \setcounter{equation}{0} 
 \renewcommand{\theequation}{B.\arabic{equation}}
 \section{Appendix B: Stability of the boundary solutions}

 Let us discuss explicitly the stability of 
 the boundary solutions appearing in  the ground state 
 and some excited states, both for the cases  $p>0$ and $p<0$. 
 We recall some notation in the following. 
 We have defined  the symbols $\kappa$ and 
  $\alpha$ by   $\kappa=|p|$ 
 and  $\alpha= \sinh \kappa /u$, respectively. 
 The symbol  $p_{cj}$ is given by 
 $p_{cj}=(j-1)u + {\sqrt{1+ (j-1)^2 u^2}}$ for some integer $j$; 
 we note that $p=p_{cj}$ corresponds to $\alpha=j-1$.

 Let us introduce a useful formula in the following 
 \begin{eqnarray}
 \theta_n (i \gamma u -v) + \theta_n (i \gamma u +v) 
 & = &  {\frac i {2 \pi }} \ln 
 \left( {\frac  {(\gamma+n)^2 u^2 + v^2} 
		{ (\gamma-n)^2 u^2 + v^2}} \right), \quad  
 {\rm for } \quad v>0, \nonumber \\
 \label{im}
 \end{eqnarray}
 where  $\gamma \ge 0$. We can show  eq. (\ref{im}) 
 by a similar method for the formula (\ref{shift}); 
 we take the  branch of the logarithmic function, and 
 use the relations 
   $arg(i(\gamma -n)u -v) = arg(v - i(\gamma -n) u) - \pi$ and  
 $arg(i(\gamma + n)u -v) =  arg(v - i (\gamma +n)) + \pi$.

 For the case of $p>0$, 
 we may consider  the three complex roots $k_1$, $k_L$ 
 and $v_1$ in the following regions. 
 \begin{eqnarray}
 k_L & =& \pi + i \kappa -i \delta_L, \qquad {\rm for} \quad \alpha > 0, 
 \nonumber \\ 
 v_1 & = & i (\alpha -1)u  + i \eta_1,  \qquad {\rm for} \quad \alpha > 1,  
 \nonumber \\
 k_{1} & = & i \log \left( (\alpha -2)u + 
 \sqrt{(\alpha - 2)^2 u^2 +1} \right) + \delta_1, 
 \quad {\rm for} \quad \alpha > 2.   
 \end{eqnarray}     
 We call the boundary solutions {\it stable} when $
 \delta_L$, $\delta_1$ and $\eta_1$ are very small. 
 For some convenience, we use  symbols $\epsilon_1$ and $\epsilon_L$   
 defined in the following 
 \begin{equation}
 \sin k_1 = i(\alpha-1)u + i \epsilon_1, \quad 
 \sin k_L = i(\alpha-1)u + i  \epsilon_L,     
 \end{equation}
 which are related to $\delta_1$ and $\delta_L$ by 
 $$
 \epsilon_L= - \cosh \kappa \times \delta_L, \qquad 
 \epsilon_1 = \sqrt{(\alpha-2)^2 u^2 +1 } \times \delta_1. 
 $$

 Let us give explicitly some
  evaluations of $\epsilon_L$, $\eta_1$ and $\epsilon_1$    
 for the case of $p>0$. We assume that 
 $p$ is not close to 
 any of the critical points $p_{cj}$'s.   
 Then for the ground state and some excited states we can show the following. 
 \begin{enumerate} 
 \item 
 When there is only one boundary solution $k_L$,  
 we have 
 \begin{equation}
 \epsilon_L = O\left(p^{-2L}\right). 
 \label{case1}
  \end{equation}
 \item 
 When there are two boundary solutions $k_L$ and 
 $v_1$ and when $u > 1$, we have 
 \begin{eqnarray}
 |\epsilon_L-\eta_1| & = & 
 O\left( \left( {\frac{\alpha}{2-\alpha}}\right)^{-2N}\right) , 
 \nonumber \\ 
 |\epsilon_L| & = & O\left( \left( {\frac{\alpha}
 {2-\alpha}}\right)^{2N} p^{-2L} \right) .
 \label{case2}
 \end{eqnarray}
 \item 
 When there are three boundary solutions $k_L$, $k_1$ and 
 $v_1$ and when $u>1$, we have 
 \begin{eqnarray}
 |\epsilon_1-\eta_1| & = & 
 O \left(|z_1|^{2L} \right) ,\nonumber \\ 
 |\epsilon_L-\eta_1| & = & 
 O\left(  |z_1|^{2L} \left( {\frac{\alpha}{\alpha-2}}\right)^{-2N}\right),
 \nonumber \\ 
 |\epsilon_L| & = & O\left( |z_1|^{-2L} 
 \left( {\frac{\alpha}{\alpha-2}} \right)^{2N} p^{-2L} \right) .
 \label{case3}
 \end{eqnarray}
 Here $z_1$ denotes the following
 $$
 z_1 = \exp(ik_1) = - (\alpha -2)u + \sqrt{(\alpha-2)^2 u^2 +1}.  
 $$
 It is easy to see that $| z_1 | < 1$ for  $\alpha > 2$. 
 \end{enumerate}
 We note that the evaluations (\ref{case1}), (\ref{case2}) and 
 (\ref{case3}) can be applied  
 for the half-filled ground-state solutions in 
 the regions of $p_{c1} < p <p_{c2}$ ($0 < \alpha < 1$), 
 $p_{c2} < p <p_{c3}$ ($1 < \alpha < 2$) , and  
 $p_{c3} < p$ ($2 < \alpha $), respectively.

 In the derivation of (\ref{case2}) and (\ref{case3}), 
 we have assumed that $u>1$. 
 In fact, 
 applying  the formula (\ref{im})
 with $\gamma = \alpha-1$, we 
 can make the following approximation when $u>1$
 \begin{eqnarray}
 \sum_j \sum_{r =\pm 1} \theta( i (\alpha-1) u -r \sin k_j) 
 & = & \sum_j {\frac i {2\pi}} \ln \left(
 {\frac {\alpha^2 u^2 + \sin^2 k_j} 
       {(\alpha-2)^2 u^2 + \sin^2 k_j} } \right) 
 \nonumber \\ 
 \approx 
 {\frac {i N} {2 \pi}} \ln \left( {\frac {\alpha^2} {(\alpha-2)^2}} \right) . 
 \label{approxm}
 \end{eqnarray}
 We recall here that $N$ is the number of electrons. 
 The approximation (\ref{approxm}) can be  not effective 
 when $u$ is very small. However,  it seems  
 that it is nontrivial to evaluate $\epsilon_L$ and $\eta_1$
 for the weak-coupling case; 
   more precise  estimates on  $\sin^2 k_j$'s should be 
 necessary when $u$ is very small.

 For the case when  (\ref{case2}) is valid,  
 the quantity $\epsilon_L$ should be very small 
 if the following inequality holds  
 \begin{equation}
 \left| p^{-1} \frac {\alpha} {2-\alpha} \right| < 1 .
 \label{ineq1}
 \end{equation}  
 For the region: $p_{c2} < p <p_{c3}$ ($1 < \alpha < 2$) , 
 however, the inequality 
 (\ref{ineq1})  does not hold for all values of $\alpha$ 
 satisfying $1 < \alpha < 2$. 
 Let us consider the case of $u\gg 1$;  
 when $u$ is very large,  we can approximate 
 $1/p$ by $1/(2\alpha u)$ using the 
 relation: $p =  \alpha u + \sqrt{1+\alpha^2 u^2}$.  
  Then, we can show  that  
 the inequality (\ref{ineq1}) holds under the following condition  
 \begin{equation}
 \alpha < 2 - {\frac 1 {2u}} . 
 \end{equation}
 Thus, at least for the case of $u \gg 1$,
 we have shown that the boundary solutions $k_L$ and $v_1$ are stable 
 when $1 < \alpha < 2-1/(2u)$, where 
 $2-1/(2u)$ is very close  to the critical point 
 $\alpha=2$. 
 For the region: $p  > p_{c3}$ ($\alpha > 2$), 
 we can  show, under the condition: $u \gg 1$,  
 that the boundary solutions $k_L$ and $v_1$ are stable 
 if $\alpha > 2 + 1/(2u)$.

 Similarly,  
 we can  discuss the case when the evaluation (\ref{case3}) 
 is valid,   
 where there are the three boundary solutions,  
 $k_L$, $k_1$ and $v_1$. 
 For the strong-coupling case, 
 we can explicitly show that 
 the boundary solutions are stable 
 if $\alpha > 2 + 1/(2u)$;  we have the following 
 \begin{equation}
 p^{-1} |z_1|^{-1} \frac {\alpha} {\alpha-2} < 1, \quad 
 {\rm when} \quad \alpha > 2 + 1/(2u). 
 \end{equation}

 Let us now consider the boundary solutions for  
 the case when $p<0$.  
 \begin{eqnarray}
 k_1^{'} & = & i \kappa -i \delta_1^{'},  
 \qquad {\rm for } \quad  0 < \alpha < 1 , 
 \nonumber \\
 v_1^{'} &  = & i (\alpha -1)u + i \eta_1^{'}, 
 \qquad {\rm for } \quad  1 < \alpha < 2 , \nonumber \\ 
 k_2^{'} & = & i \log\left( (\alpha -2)u + {\sqrt{(\alpha-2)^2 u^2 +1} }
 \right) 
 +i \delta_2^{'},  
 \, {\rm for } \, 2 <  \alpha  . 
 \end{eqnarray}
 Here $\delta_1^{'}$, $\delta_2^{'}$ and $\eta_1^{'}$ should be very small. 
 We recall that for the case of $p<0$,  the boundary solutions 
 $k_1^{'}, k_2^{'}$ and $v_1^{'}$ 
 have been discussed in Ref. \cite{BF}. 
 For some convenience, we use  symbols $\epsilon_1^{'}$ and 
 $\epsilon_2^{'}$ 
 defined in the following 
 \begin{equation}
 \sin k_1^{'} = i(\alpha-1)u + i \epsilon_1^{'}, \quad 
 \sin k_L^{'} = i(\alpha-1)u + i  \epsilon_2^{'},    
 \end{equation}
 which are  related to $\delta_1^{'}$ and $\delta_2^{'}$ 
 by 
 $$
 \epsilon_1^{'}= - \cosh \kappa \times \delta_1^{'}, \qquad 
 \epsilon_2^{'} = \sqrt{(\alpha-2)^2 u^2 +1 } \times \delta_2^{'}. 
 $$

 Applying the formula (\ref{im}), 
 we can evaluate $\epsilon_1^{'}$, $\eta_1^{'}$ 
 and $\epsilon_2^{'}$ as follows. 
 \begin{enumerate} 
 \item 
 When there is only one boundary solution $k_1^{'}$, 
 we have 
 \begin{equation}
 \epsilon_1^{'} = O \left( p^{-2L} \right). 
 \label{case-1} 
 \end{equation}
 \item 
 When there are two boundary solutions $k_1^{'}$ and $v_1^{'}$ 
 and when $u>1$, 
 we have 
 \begin{eqnarray}
 |\epsilon_1^{'}-\eta_1^{'}| & = & 
 O \left( \left( {\frac{\alpha}{2-\alpha}}\right)^{-2N} \right) ,
 \nonumber \\ 
 |\epsilon_1^{'}| & = & O \left( \left( {\frac{\alpha}
 {2-\alpha}}\right)^{-2N} p^{-2L} \right)  . 
 \label{case-2}
 \end{eqnarray}
 \item 
 When there are three boundary solutions $k_1^{'}$, $k_2^{'}$ and $v_1^{'}$ 
 and when $u>1$, we have 
 \begin{eqnarray}
 |\epsilon_2^{'}-\eta_1^{'}| & = & 
 O \left( |z_2^{'}|^{2L} \right) , \nonumber \\ 
 |\epsilon_1^{'}-\eta_1^{'}| & = & 
 O \left(  |z_2^{'}|^{2L} \left( {\frac{\alpha}{\alpha-2}}\right)^{-2N} 
\right) , 
 \nonumber \\ 
 |\epsilon_1^{'}| & = & O \left( |z_2^{'}|^{2L} 
 \left( {\frac{\alpha}{\alpha-2}} \right)^{-2N} p^{-2L} \right) . 
 \label{case-3}
 \end{eqnarray}
 Here $z_2^{'}$ is given by 
 $$
 z_2^{'}= \exp(ik_2^{'}) = -(\alpha -2)u + \sqrt{(\alpha-2)^2u^2 +1 } . 
 $$
 We note that $|z_2|<1$ when $\alpha >2$. 
 \end{enumerate}
 We note that the evaluations (\ref{case-1}), 
 (\ref{case-2}) and (\ref{case-3}) can be applied 
 for the ground-state solutions in the regions of  
 $-p_{c2} < p < -p_{c1}$ ($0 < \alpha < 1$),  
 $-p_{c3} < p < -p_{c2}$ ($1 < \alpha < 2$),  
 and $p <- p_{c3}$ ($2 < \alpha $).  
 We also note that in the derivations of 
 (\ref{case-2}) and (\ref{case-3}) we have made the same approximation
 with (\ref{approxm}).

 For the case when the evaluations (\ref{case-1}), 
 (\ref{case-2}) and (\ref{case-3}) 
  are valid, 
 the stability conditions for the 
 boundary roots are satisfied for any  $p$ and $u$.   
 For the case we have  the following relations 
 when $p<-1$ and $u>0$  
 \begin{equation}
 |p|^{-1} < 1, \quad 
 |p^{-1} \frac {\alpha} {\alpha-2}| < 1, \quad 
 |p^{-1} z_2^{'} \frac {\alpha} {\alpha-2}| < 1, \, etc. 
 \end{equation}
 However, we should remark that it is not certain 
 whether  (\ref{case-2}) and (\ref{case-3}) are valid 
 also for the weak-coupling case: $u \ll 1$.

 We now discuss how 
  the boundary solutions of the 
 open-boundary Hubbard model can be related to those  
 of the interacting spin-1/2 fermion systems. We  consider 
 the case when the band width $t$ 
 is very large  and the electron density $N/L$ is very small.

 In order to show explicitly the effect of the large band-width,  
 we  replace $u$ and $p$ in eqs. (\ref{BAE}), (\ref{ZLcs}) and 
 (\ref{zBcs}) by 
 $u/t$ and $p/t$, respectively. 
 We recall that so far the energy scale has been 
  normalized such that $t=1$. 
 Under the limit of  $t \rightarrow \infty$,  
 the critical points $p_c's$ become 
  the following  
 \begin{eqnarray}
 p_{c1}/t & \rightarrow &  1,  \nonumber \\
 p_{c2}/t & \rightarrow & 1+ u/t ,  \nonumber \\
 p_{c3}/t & \rightarrow & 1+ 2u/t.    
 \end{eqnarray}
 The values obtained in the limit
  are equivalent  to the critical points 
 of the boundary parameter \cite{spin-1/2} 
 for the interacting spin-1/2 fermions.

 When the electron density $N/L$ is very small, the Fermi wavenumber 
 $k_F$ should  be  very small. Therefore,  
 we can make linear approximations for the charge rapidities such as 
 $\exp(ik) \approx 1 + ik$ and $\sin k \approx k$. 
 Then, we can show that the boundary term in the Bethe ansatz equations 
 of the open-boundary Hubbard model corresponds to that of the interacting 
 spin-1/2 fermion system under the linear  approximations   
 \begin{equation}
 {\frac {2k} {2 \pi}} - 
 {\frac 1 {2 \pi i}} 
 \ln \left( {\frac {1 + \exp(ik)p/t} 
		   {1 + \exp(-ik)p/t} } \right) 
 \rightarrow 
 - {\frac 2 { 2 \pi} } \tan^{-1} \left( {\frac k {-(1+p/t)} } \right) . 
 \end{equation}
 Hereafter we may renormalize the boundary chemical potential $p$ so 
 that we can replace  $p/t$  by $p$. 
 Thus, we have  explicitly shown that when $t \gg 1$ and $N/L \ll 1$, 
 the Bethe ansatz equations of the open-boundary 1D Hubbard model 
 are reduced into those  
 of the interacting spin-1/2 fermion system \cite{spin-1/2} 
 with the open-boundary condition.

 Under the large band-width and small electron-density limit, 
 the boundary solutions 
 of the open-boundary Hubbard model for the case 
 of $p<0$ remain intact. We can make the same approximation 
 with (\ref{approxm}), since $\sin^2 k_j$'s are very small. 
 Here we note that the case of large band-width corresponds to  
 the weak-coupling case where the approximation (\ref{approxm}) 
 can be non-effective. When the density is very low, however, 
 it is  valid for some cases.  
 For example, 
 we may consider the case where $N$ is fixed and $L$ 
 is proportional to 
 $t$ under the limit $t \rightarrow \infty$  
 so that each momentum $k_j$ is proportional to $1/t$. 
 Then, we can apply the approximation (\ref{approxm}) for this case.    
 In this way, the boundary solutions of the open Hubbard model 
 for the case $p<0$
  are related to those   of the interacting 
 spin-1/2 fermion system discussed in Ref. \cite{spin-1/2}.

 For the case when $p>0$, however, 
 the boundary solutions of the open Hubbard model 
 are not related to any solution of the interacting spin-1/2 
 fermion system. 
 They exist only when the band is half-filled. 
 The physical condition is completely different from the 
 low density case.

 \setcounter{equation}{0} 
 \renewcommand{\theequation}{C.\arabic{equation}}

 \section{Appendix C: Particle-hole transformation 
 for the open-boundary Hubbard chain}

 Let us denote 
 by $d_{j,\sigma}$ and  $d_{j,\sigma}^{\dagger}$,
 the annihilation and creation operators  
 for the hole with spin $\sigma$ on the $j$th site, 
  respectively.  We define a particle-hole transformation 
 by  the following. 
 We replace 
 the creation (annihilation) operator 
 of electron with spin $\sigma$ on the $j$th site 
  by the annihilation (creation) operator 
 of  hole with spin $\sigma$ on the $j$th site   
 for $\sigma= \uparrow, \downarrow$ 
 and for $j=1,\ldots,L$, and then multiplying 
 the gauge factor $(-1)^j$ to the hole operators on the $j$th site 
 for over all the sites:  
 \begin{equation}
 c_{j,\sigma}^{\dagger} \rightarrow (-1)^jd_{j,\sigma}, \qquad  
 c_{j,\sigma}
  \rightarrow (-1)^j d_{j,\sigma}^{\dagger} . 
 \label{pht}
 \end{equation}

 The ground-state energy for $p>0$ is related to that of $p<0$ 
 by the particle-hole transformation; the sign of the boundary chemical 
 potential is changed under the transformation. 
 Let us denote by 
 $E(N_{\downarrow},N_{\uparrow}; U, p)$ 
 the ground-state energy for $N_{\downarrow}$ down-spin electrons, 
 $N_{\uparrow}$ up-spin electrons 
 with the Hubbard coupling $U$ and the boundary chemical potential $p$. 
 Then, applying the particle-hole transformation , we have the following 
 \begin{equation}
 E(L-M, L-M^{'}; U,p) = E(M,M^{'}; U, -p)+(L-N)U + 2p . 
 \end{equation}

 For the half-filled band, 
 the ground-state energies for $p>0$ and $p<0$ are explicitly related.  
 Recall we assume $L$ is even. Then, we have the following 
 \begin{equation}
 E(L/2, L/2; U,p) = E(L/2, L/2; U, -p) + 2p . 
 \end{equation}

 On the other hand, it seems quite difficult to find out 
 an  explicit relation between  the sets of  
  the charge (spin) rapidities for the cases $p>0$ and $p<0$. 
 It seems as if there might be such a simple connection 
 that for any   momentum $k$ in the ground state of $p>0$  
 the value $\pi \pm k$ corresponds to one of the ground-state solutions   
 for $p<0$.  However, it is not the case. 
 There is no such relation between 
 the boundary solutions $k_1^{'}$, $k_2^{'}$ and $v_1^{'}$ for 
 $p <-p_{c3}$ and the boundary solutions 
 $k_1$, $k_L$ and $v_1$ for $p>p_{c3}$. 

 From some numerical solutions of the Bethe ansatz equations, 
 it is suggested that it can be quite nontrivial to 
 find out any  explicit relations between 
 the sets of the half-filled 
 ground-state solutions for the cases $p>0$ and $p<0$. 
 Some details should be discussed in later papers.

\setcounter{equation}{0} 
\renewcommand{\theequation}{D.\arabic{equation}}

\section{Appendix D: Boundary solutions for some excited states}

We discuss the quantum numbers of some excited states with 
the boundary solutions. We assume the adiabatic hypothesis for 
the quantum numbers.

Let us consider  an excited state which have 
only real-valued momenta and rapidities 
when $p=0$. We denote 
by  $\Delta^c_0$  ( $\Delta^s_0$) the set of 
the quantum numbers for the  momenta ( rapidities).  
Let us  denote by $\Delta^c (\Delta^s)$  the set of 
the quantum numbers of  real-valued momenta in the excited state 
at a given value of $p$.  
In general,  $\Delta^c (\Delta^s)$  depends on $p$. 
For the excited state 
we write by $\Delta_{im}^c$ ($\Delta_{im}^s$) the set of 
the quantum numbers  of complex-valued momenta (rapidities).  
It is useful to introduce 
the notation for holes; we denote 
by $\Delta_{hole}^c$ ($\Delta_{hole}^s$) 
the set of the quantum numbers of  
holes for the real-valued momenta (rapidities). 
Then, we can 
define $Z_L^c(k)$, $Z_L^s(v)$, $z_B^c(k)$, and $z_B^s(v)$ also for 
the excited state; in the formulas (\ref{ZLcs}) 
we replace $\Delta_g^c$ and $\Delta_g^s$ 
by $\Delta^c$ and $\Delta^s$, respectively. 
Similarly to the ground state, 
 we can evaluate  the $I_{min}$'s for the excited state 
as follows. 
\begin{eqnarray}
I_{min} & = & z_B^c(0) +1, \qquad 
I_{max}  =   L+ z_B^c(\pi) -1, \nonumber \\
J_{min} & = & z_B^s(0) +1, \nonumber \\ 
J_{max} & = & (N-N_{im})-(M-M_{im})+ (z_B^s(\infty) -1/2) 
\label{minmax} . 
\end{eqnarray}
Here $N_{im}$ and $M_{im}$ denote the number of 
complex-valued charge and spin rapidities 
(boundary solutions), respectively. 
We recall that 
$N$ and $M$ denote the number of 
electrons and that of down-spins, respectively.

For an illustration, let us discuss the boundary solutions 
of an excited state for the case when $p<0$. 
Hereafter we assume $N<L$. 
We consider  the excited state of 
$N$ electrons with $M$ down-spins 
where the quantum number  at $p=0$ is given by the following. 
\begin{eqnarray}
\Delta^c_0 & = & \{1,  3, 4, \ldots, N+1 \},  
\qquad 
\Delta^s_0  =  \{1, 2, \ldots, M \},  
\nonumber \\ 
\Delta_{im}^c  & = & \quad \Delta_{im}^s= \phi .
\end{eqnarray}
It follows from (\ref{minmax}) 
that when $p=0$ the sets of holes are given by 
\begin{equation}
\Delta_{hole}^c   =  \{ 2, N+2,N+3, \ldots, L \},  
\quad \Delta_{hole }^s= \{M+1, \ldots, N-M \} .
\end{equation}
Here we note that when 
$L=N+1$, then we have $\Delta_{hole}^c= \{2 \}$, and also that 
when  $N$ is even and $M=N/2$, then we have $\Delta_{hole}^s = \phi$. 
Applying the formulas (\ref{shift}) and 
 (\ref{minmax}), we can  show 
that there are four critical points given by 
$-p_{cj}$ for $j = 1, \ldots, 4$. We have the following 
five cases when $p<0$. 
\begin{enumerate}
%
\item For $- p_{c1}< p < 0 $,  
we have no boundary solution. We have a hole at $I=2$.
$$
\Delta_{im}^c = \Delta_{im}^s=\phi, 
$$
$$
\Delta^c  =  \{1,  3, 4, \ldots, N+1 \}, \quad 
\Delta^s  =  \{1, 2, \ldots, M \} ,  
$$
$$
\Delta_{hole}^c=\{ 2, N+2, \ldots, L \}, \quad 
\Delta_{hole}^s= \{ M+1, \ldots, N-M \} . 
$$
%
\item  For $-p_{c2}< p< - p_{c1}$, 
 we have $k_1^{'}$ and a hole at $I=2$. 
$$
 \Delta_{im}^c=\{ 1 \}, \qquad 
 \Delta_{im}^s=\phi, 
$$
$$
\Delta^c  =  \{ 3, 4, \ldots, N+1 \}, \quad 
\Delta^s  =  \{ 2, 3, \ldots, M \} , 
$$
$$
 \Delta_{hole}^c=\{ 2, N+2, \ldots, L \}, \quad 
\Delta_{hole}^s=\{ M+1, \dots, N-M  \} . 
$$
%
\item For $-p_{c3} < p< -p_{c2}$,  
 we have $k_1^{'}$ and $v_1^{'}$ and a hole at $I=2$. 
$$
 \Delta_{im}^c= \{ 1 \}, \qquad 
 \Delta_{im}^s= \{ 1 \}, 
$$
$$
\Delta^c  =  \{  3, 4, \ldots, N+1 \} , \quad
\Delta^s  =  \{ 2, 3, \ldots, M \} , 
$$
$$
 \Delta_{hole}^c= \{ 2, N+2, \ldots, L \}, \quad 
 \Delta_{hole}^s= \{M+1, \dots, N-M \}  . 
$$ 
\item For $-p_{c4} < p < -p_{c3}$,  
we have $k_1^{'}$ and $v_1^{'}$ but no hole at $I=2$. 
A new hole appears at $I=L+1$. 
$$
 \Delta_{im}^c=\{1 \}, \qquad 
 \Delta_{im}^s=\{ 1 \}, 
$$
$$
\Delta^c  =  \{ 3, 4, \ldots, N+1 \}, \quad 
\Delta^s  =  \{ 2, 3, \ldots, M \} , 
$$
$$
\Delta_{hole}^c= \{N+2, \ldots, L+1 \}, 
\quad 
 \Delta_{hole}^s= \{M+1, \ldots, N-M \} . 
$$ 
\item  For $ p< - p_{c4}$, 
we have $k_1^{'}$ and $v_1^{'}$. 
A new hole appears at $J=1$. 
$$
 \Delta_{im}^c=\{ 1 \}, \qquad 
 \Delta_{im}^s=\{ 1 \},
$$
$$
 \Delta^c  =  \{3, 4, \ldots, N+1 \}, \quad 
\Delta^s  =  \{ 2, 3, \ldots, M \} , 
$$
$$
 \Delta_{hole}^c= \{N+2, \ldots, L+1 \}, \quad 
 \Delta_{hole}^s=\{1,M+1,\ldots, N-M  \}  . 
$$
We recall $p_{c4}  =  3u + \sqrt{1+ (3u)^2}$ . 
\end{enumerate}

Let us consider the ground state and the excited state 
discussed in the last paragraph.  
For the two regions $-p_{c4} < p < -p_{c3}$ and $p< -p_{c4}$,   
the ground-state solutions have the same structure,  
while the excited state  have the different structures; 
the excited state has the two 
 boundary solutions both for the two regions, 
 however, it has  the different numbers of holes in the spin rapidities 
 for the two regions.  Thus, it is 
 suggested that there can be  more subtle points in  
 the boundary excitations than had been described  
in Ref. \cite{spin-1/2} for those of 
the interacting spin-1/2 fermion system. 
However, it seems that some physical interpretations  
similar to those in Ref. \cite{spin-1/2} should be valid 
also for the boundary excitations of the open-boundary Hubbard model.
Some precise investigations  
should be discussed in later publications.

\end{document}